\documentstyle[aps,floats,psfig]{revtex}
\newcommand{\nad}[1]{\underline{\vec#1}}

\begin{document}
\title{Decays of spacelike neutrinos\thanks{This work is supported by 
University of {\L}\'od\'z grant No.~488.}}
\author{\rm Pawe{\l} Caban\thanks{caban@mvii.uni.lodz.pl}
\and Jakub Rembieli\'nski\thanks{jaremb@mvii.uni.lodz.pl and 
jaremb@krysia.uni.lodz.pl}
\and Kordian A. Smoli\'nski\thanks{xmolin@mvii.uni.lodz.pl}}
\address{Department of Theoretical Physics, University of {\L}\'od\'z\\
ul.~Pomorska 149/153, 90--236 {\L}\'od\'z, Poland}
\wideabs{
\maketitle
\begin{abstract}
In this paper we consider the hypothesis that neutrinos are fermionic 
tachyons with helicity $1/2$. We propose an effective model of interactions 
of these neutrinos and analyze dominant effects under the above hypothesis: 
the three-body neutrino decay, the radiative decay and briefly discuss 
$\beta$-decay with tachyonic antineutrino. We calculate the corresponding 
amplitudes and the mean life time for decays, as well as the 
differential decay width. In addition we apply the neutrino three-body 
decay to the solar neutrino problem and we obtain a remarkable change of 
energy spectrum of the neutrino flux at the Earth ground (assuming no 
neutrino oscillation). We compare the results with the Standard Solar Model 
predictions and solar neutrino experiments to estimate an upper bound for the
taonic neutrino.
\end{abstract}
}

\narrowtext
\footnotetext[1]{This work is supported by University of {\L}\'od\'z grant
No.~488.}
\footnotetext[2]{caban@mvii.uni.lodz.pl}
\footnotetext[3]{jaremb@mvii.uni.lodz.pl and 
jaremb@krysia.uni.lodz.pl}
\footnotetext[4]{xmolin@mvii.uni.lodz.pl}

\section{Introduction}\label{sec:1}
The indication for negative square of electron and muon neutrinos 
four-momenta in all the recent experiments \cite{Bar96,Bel95} requires a 
theoretical or methodological explanation. However, up to now, there is no 
satisfactory way to understand this phenomenon. The most popular solution of 
this problem is a possible presence of molecular effects in the final state; 
however they are not understood well \cite{Ell96}. Another proposed 
possibility is an existence of a hidden anomalous long range interaction of 
neutrinos \cite{Moh96} or a presence of a repulsive potential originating in 
coherent interaction of neutrino with the daughter atom \cite{Col96}. Also 
some methodological misinterpretations can cause the problem, as it was 
suggested by Roos and Khalfin \cite{Roo96}; see also \cite{Kap97} for a
discussion of the negative $m_\nu^2$ puzzle. However, if the negative squared 
antineutrino mass in tritium beta decay is genuine, then it is hard to 
understand this phenomenon using conventional particle physics ideas. 
Moreover an anomalous electron/muon composition of flavor of atmospheric 
neutrino flux \cite{Ell96,Gai95} as well as the solar neutrino problem is 
a puzzle, which is not solved to the end, even if we take into account 
possibility of flavor oscillations and Mikheyev--Smirnov--Wolfenstein (MSW)
mechanism \cite{Smi96,Kru96,Bil96a,Bil96b}. 

In this paper we continue investigation of an unconventional possibility that 
neutrinos are fermionic tachyons. This hypothesis was proposed firstly by 
Chodos et al.\ \cite{Cho85}. In the papers by Rembieli\'nski 
\cite{Rem96} a causal classical and quantum theory of tachyons was elaborated 
and next applied by Ciborowski and Rembieli\'nski to explanation of the 
results of tritium $\beta$-decay experiments \cite{Cib96a,Cib96b}. In the 
paper \cite{Cib96a} physical consequences of the three body decay $\nu_\ell 
\to \nu_\ell \bar{\nu}_{\ell'} \nu_{\ell'}$ was also discussed. This paper is 
devoted to systematical calculations of decay rates for the two processes 
with participation of neutrinos: the radiative decay $\nu_\ell \to 
\nu_\ell \gamma$ and the three body decay $\nu_\ell \to \nu_\ell
\bar\nu_{\ell'} 
\nu_{\ell'}$, both conserving the lepton flavor; we briefly discuss also the 
$\beta$-decay. Notice, that the first two reactions are kinematically 
admissible for tachyonic neutrinos only. It is remarkable that, as was 
stressed in \cite{Cib96a}, the emission by neutrino $\nu_\ell$ a 
$\bar{\nu}_{\ell'} \nu_{\ell'}$ pair can be an additional, qualitatively 
similar process to flavor oscillations. As we will see (Sec.~\ref{sec:5} and 
\ref{sec:6}), the calculated mean times of life for the aforementioned decays 
take reasonable values in the regions of experimentally available energies. 
Notice however that applicability of these results to the description of the 
time evolution of a neutrino flux is restricted at this level because both 
of the decays form cascades. Therefore they demand further investigation as 
nontrivial stochastic processes leading to different than Geiger--Nutall decay 
law. Moreover, the mostly hopefull program is to involve both oscillations 
and three body decay. The corresponding calculations for solar and 
atmospheric neutrinos as well as for the supernova SN1987A neutrino flux 
are now in progress. 
However we give here an estimation for the upper bound of the 
tau neutrino mass (if it has tachyonic character) by means of the solar 
neutrino data.

\section{Preliminaries}\label{sec:2}
In this chapter we give a brief presentation of the causal theory of tachyons 
proposed in the paper \cite{Rem96}.

It is a common opinion that existence of space-like particles is in 
contradiction with special relativity. This is because of the 
Einstein--Poincar\'e (EP) causality violation and consequently because of a 
number of very serious difficulties like causal paradoxes, inconsistent 
kinematics with impossibility of a covariant formulation of the Cauchy 
problem, unbounded from below energy spectrum and many others. Even more 
problems arise on the quantum level, like the problems with construction of 
covariant asymptotic spaces of states, vacuum instability, etc.

However, as was shown in the paper \cite{Rem96} it is possible to agree 
special relativity with the tachyon concept. The main idea is based on the 
well known fact that the definition of the coordinate time depends on the 
synchronization scheme \cite{Jam79,Man77,Wil93} which, in turn, is a 
convention related to the assumed one-way light velocity (only the average of 
the light velocity over closed paths has an operational meaning; in special 
relativity this average is frame independent). Taking into account this 
freedom, it is possible to realize Poincar\'e group transformations in such a 
way that the constant-time hyperplane is a covariant notion \cite{Rem96}. In 
that synchronization (named Chang--Tangherlini synchronization scheme and 
denoted as CT synchronization---see \cite{Rem96} and references therein) 
it is possible to overcome all the 
difficulties which appear in the standard approach.

Let us stress main features of this nonstandard scheme. Firstly, {\em it is 
fully equivalent to the standard formulation of the special relativity if we 
restrict ourselves to the timelike and lightlike trajectories}, i.e. if we 
exclude tachyons. If we admit space-like trajectories (tachyons are included) 
then one inertial frame is distinguished as a preferred frame\footnote{We 
would stress that the notion of preferred frame used through this work is not 
the one used in the search of breaking the Lorentz symmetry.}. Thus {\em the 
relativity principle is broken, however the Poincar\'e symmetry is still 
preserved in that case}. The proper framework to this construction is the 
bundle of Lorentzian frames; the base space is simply the space of velocities 
of these frames with respect to the preferred frame. For this reason the 
transformation law for coordinates involves velocity of distinguished 
frame. The preferred frame can be locally identified with the comoving frame 
in the expanding universe (cosmic background radiation frame), i.e. the 
reference frame of the privileged observers to whom the universe appears 
isotropic \cite{Wei72}.

Now, the velocity of the Solar System deduced from the dipole anisotropy of 
the background radiation is about $350$~km/sec \cite{lubin83}, so it is almost
at 
rest relatively to the preferred frame. Therefore, with a good approximation, 
we can perform calculations in the preferred frame.

To be concrete, the Lorentz group transformations in the mentioned bundle of 
frames have the following form \cite{Rem96}
\begin{equation}
\begin{array}{l}\displaystyle
x' = D(\Lambda,u) x\,, \\ \displaystyle
u' = D(\Lambda,u) u\,,
\end{array}\label{p1}
\end{equation}
where for rotations $D(R, u)$ has the standard form while for boosts it reads
\begin{eqnarray}\label{D:Wu}
\lefteqn{D(W, u) }\nonumber\\
&&= \left(\begin{array}{c|c} \left(W^0\right)^{-1} & 0 \\ 
\hline
-\vec{W} & I +\frac{\vec{W} \otimes \vec{W}^T}
{1 + \sqrt{1 + (\vec{W})^2}}
-\vec{W} \otimes \vec{u}^T u^0 \end{array}\right)\,.
\end{eqnarray}
Here $W^{\mu}$ is the four-velocity of $(x')$ frame seen by an observer in 
the frame $(x)$ while $u^{\mu}$ is the four-velocity of the privileged frame 
as seen from the frame $(x)$. Notice that the time coordinate is rescalled by 
a positive factor only. The transformations (\ref{p1}) leave invariant the 
metric form 
\begin{equation}
ds^2 = g_{\mu\nu}(u) dx^{\mu} dx^{\nu}  \label{p3}
\end{equation}
with
\begin{equation}\label{g:u}
g(u) = \left(\begin{array}{c|c} 1 & u^{0}
\vec{u}^T \\ 
\hline
u^{0} \vec{u} & -I + \vec{u} \otimes \vec{u}^T (u^{0})^{2}
\end{array}\right)\,.
\end{equation}
Interrelation with coordinates in the Einstein--Poincar\'e synchronization 
$(x_{\rm E}^{\mu})$ is given by 
\begin{equation}
x_{\rm E}^{0} = x^0 + u^0 \vec{u}\cdot\vec{x}\,,\quad \vec{x}_{\rm
E}=\vec{x}\,.
\label{p4}
\end{equation}
However the corresponding interrelations between velocities $\vec{v}_{\rm E}$ 
and $\vec{v}$ obtained from (\ref{p4}) are singular for superluminal 
velocities.

Now, on the quantum level, the corresponding quantum mechanics and quantum 
field theory can be formulated with help of the bundle of Hilbert spaces 
associated with the aforementioned bundle of frames. The Fock construction 
can be done in this framework and unitary orbits can be classified 
\cite{Rem96}. For usual particles they coincide with the standard unitary 
representations of the Poincar\'e group. For tachyons they are induced from 
$SO(2)$ group (instead of $SO(2,1)$) and labelled by the helicity. The 
following two facts, true only in CT synchronization, are extremely important 
for quantization of tachyons:
\begin{itemize}
\item Invariance of the sign of the time component of the space-like
four-mo\-men\-tum, i.e. $\epsilon(k^0)={\rm inv}$, which follows from
the transformation law (\ref{p1})--(\ref{D:Wu}).
\item Existence of a covariant lower energy bound; in terms of the
contravariant space-like four-mo\-men\-tum $k^{\mu}$
$(k^{2}<0)$
this lower
bound is exactly zero, i.e. $k^{0}\geq0$ as in the timelike and
lightlike case.
\end{itemize}
This is the reason why an invariant Fock construction with a stable vacuum 
can be done in this case on the quantum level \cite{Rem96}. Notice that the 
measure
\begin{equation}\label{dmu}
d\mu(k, \kappa) = d^4k\, \theta(k^0)\, \delta(k^2 + \kappa^2)
\end{equation}
is Poincar\'e covariant in the CT synchronization scheme.

\section{Fermionic tachyons with helicity $\lambda=\pm1/2$}\label{sec:3}
To construct tachyonic field theory describing field excitations with the 
helicity $\pm1/2$, we assume that our field transforms under Poincar\'e group 
like bispinor (for discussion of transformation rules for local fields in the 
CT synchronization see \cite{Rem96,Rem80}); namely
\begin{equation}
U(\Lambda) \psi(x, u) U(\Lambda^{-1}) = S(\Lambda^{-1}) \psi(x', u')\,,
\end{equation}
where $S(\Lambda)$ belongs to the representation $D^{\frac{1}{2}0} \oplus 
D^{0\frac{1}{2}}$ of the Lorentz group. Because we are working in the CT 
synchronization, it is convenient to introduce an appropriate (CT-covariant) 
base in the algebra of Dirac matrices as
\begin{equation}
\gamma^{\mu}={T(u)^\mu}_{\nu}\gamma^{\nu}_{\rm E}\,,
\end{equation}
where $\gamma^{\mu}_{\rm E}$ are standard $\gamma$-matrices, while $T(u)$ is 
determined by the Eq.~(\ref{p4}). Therefore
\begin{equation}
\{\gamma^{\mu},\gamma^{\nu}\}=2g^{\mu\nu}(u)I\,.
\end{equation}
However, notice that the Dirac conjugate bispinor $\bar\psi = \psi^\dagger 
\gamma^0_{\rm E}$. Furthermore $\gamma^5 = -i 
\varepsilon_{\mu\nu\sigma\lambda} \gamma^\mu \gamma^\nu \gamma^\sigma 
\gamma^\lambda/4! = \gamma^5_{\rm E}$. Let us consider the Dirac-like 
equation\footnote{Hereafter $u\gamma = u_\mu\gamma^\mu$, 
$u\partial=u^\mu\partial_\mu$, $\gamma\partial=\gamma^\mu\partial_\mu$.} 
\cite{Rem96}
\begin{equation}\label{dirac}
\left(\gamma^5(i\gamma\partial) - \kappa\right)\psi = 0\,.
\end{equation}
which can be derived from the Lagrangian density
\begin{equation}
{\cal L} = \bar{\psi} \left(\gamma^5(i\gamma\partial) - \kappa\right)\psi\,.
\label{lag}
\end{equation}
Dirac equation (\ref{dirac}) imply the Klein--Gordon equation
\begin{equation}
\left(g^{\mu\nu}(u) \partial_{\mu} \partial_{\nu} - \kappa^2\right) \psi = 
0\,,
\end{equation}
related to the space-like dispersion relation $k^{2} = -\kappa^{2}$. The 
equation (\ref{dirac}) is analogous to the one by Chodos et al.\ 
\cite{Cho85} Dirac-like  equation for tachyonic fermion. However, contrary to 
the standard EP approach,  it can be consistently quantized in the CT scheme 
if it is supplemented by the covariant helicity condition \cite{Rem96}
\begin{equation}
\hat{\lambda}(u) \psi(u, k) = \lambda \psi(u, k) \label{lambda}
\end{equation}
with $\hat{\lambda}$ given by 
\begin{equation}\label{lam}
\hat{\lambda}(u) = -\frac{\hat{W}^{\mu} u_{\mu}}{\sqrt{(P u)^2 - P^2}}\,,
\end{equation}
where
\begin{displaymath}
\hat{W}^{\mu} =  1/2\; \varepsilon^{\mu\sigma\lambda\tau} J_{\sigma\lambda} 
P_{\tau}
\end{displaymath}
is the Pauli--Lubanski four-vector. This condition is quite analogous to the 
condition for the left (right) bispinor in the Weyl's theory of the massless 
field. It implies that particles described by $\psi$ have helicity 
$-\lambda$, while antiparticles have helicity $\lambda$. For the obvious 
reason in the  following we will concentrate on the case $\lambda = 1/2$.

Notice that the pair of equations (\ref{dirac}), (\ref{lambda}) is not
invariant 
under the $P$ or $C$ inversions separately.

Now, in the bispinor realization the helicity operator $\hat\lambda$ has the 
following explicit form	\cite{Rem96}
\begin{equation}
\hat\lambda(u) = \frac{\gamma^5 [i \gamma \partial, u \gamma]}
{4\sqrt{(i u \partial)^2 + \Box}} \label{***}
\end{equation}
where the integral operator 
$\left((i u \partial)^2 + \Box\right)^{-1/2}$ in the coordinate 
representation is given by the well behaving distribution
\begin{eqnarray}
\lefteqn{\left((-i u \partial)^2 + \Box\right)^{-1/2}} \nonumber\\ 
&=& (2 \pi)^{-4} \int d^4p\, \epsilon(u p) e^{i p x} \left((u p)^2 - 
p^2\right)^{-1/2}\,.
\label{****}
\end{eqnarray}

The free field $\psi(x,u)$ has the following Fourier decomposition
\begin{eqnarray}
\lefteqn{\psi(x, u) = (2\pi)^{-3/2} \int d^4k\, \delta(k^2 + \kappa^2)
\theta(k^0)} \nonumber\\
&\times&\left[w(k,u) e^{i k x} b^\dagger(k) + v(k,u) e^{-i k x} 
a(k)\right]\,. \label{*****}
\end{eqnarray}
The creation and annihilation operators of a tachyonic fermion $(a)$ with 
helicity $-1/2$ and an antifermion $(b)$ with helicity $1/2$ satisfy almost 
standard canonical anticommutation relations; the nonzero ones are
\begin{eqnarray}\label{a1}
\left[a(k), a^{\dagger}(p)\right]_+ &=& 2 \omega_{\vec{k}} 
\delta(\nad{k} - \nad{p})\,, \\
\label{a2}
\left[b(k), b^{\dagger}(p)\right]_+ &=& 2 \omega_{\vec{k}} 
\delta(\nad{k} - \nad{p})\,.
\end{eqnarray}
Here $\omega_{\vec{k}} = k^0 > 0$ is the positive root of the dispersion 
relation $k^2 = -\kappa^2$ and $\nad{k}$ denotes covariant components $k_i$, 
$i = 1, 2, 3$ of $k$. The amplitudes $w$ and $v$ satisfy
\begin{eqnarray}\label{E1}
\left(\kappa + \gamma^5 k \gamma\right) w(k,u) &=& 0\,, \\
\label{E2}
\left(1 + \frac{\gamma^5 [k \gamma, u \gamma]}{2 \sqrt{q^2 + 
\kappa^2}}\right) w(k,u) &=& 0\,, \\
\label{E3}
\left(\kappa - \gamma^5 k \gamma\right) v(k,u) &=& 0\,, \\
\label{E4}
\left(1 + \frac{\gamma^5 [k \gamma, u \gamma]}{2 \sqrt{q^2 + 
\kappa^2}}\right) v(k,u) &=& 0\,.
\end{eqnarray}
Furthermore, the projectors $w \bar{w}$ and $v \bar{v}$ read
\begin{eqnarray}\label{w-w}
w(k, u) \bar{w}(k, u) &=&  \frac{\kappa - \gamma^5 k \gamma}{2} 
\left(1 - \frac{\gamma^5 [k \gamma, u \gamma]}{2 \sqrt{q^2 + 
\kappa^2}}\right)\,, \\
\label{v-v}
v(k, u) \bar{v}(k, u) &=&  -\frac{\kappa + \gamma^5 k \gamma}{2} 
\left(1 - \frac{\gamma^5 [k \gamma, u \gamma]}{2 \sqrt{q^2 + 
\kappa^2}}\right)\,.
\end{eqnarray}
The above amplitudes fulfill the covariant normalization conditions
\begin{eqnarray}\label{N1}
\bar{w}(k,u) \gamma^5 u \gamma w(k,u) = \bar{v}(k,u) \gamma^5 u 
\gamma v(k,u) &=& 2uk\,, \\
\label{N2}
\bar{w}(k^\pi,u) \gamma^5 u \gamma v(k,u) &=& 0\,,
\end{eqnarray}
where the superscript $\pi$ denotes the space inversion of $k$.

It is easy to see that in the massless limit $\kappa \to 0$ the 
Eqs.~(\ref{E1})--(\ref{E4}) give the Weyl equations
\[ k \gamma w = k \gamma v = 0\,, \quad \gamma^5 w = -w\,, \quad \gamma^5 v = 
-v\,. \]

Now, the normalization conditions (\ref{N1})--(\ref{N2}) together with the 
canonical commutation relations (\ref{a1}), (\ref{a2})  guarantee the proper 
work of the canonical formalism. In particular, starting from the Lagrangian 
density (\ref{lag}) we can  derive the translation generators
\begin{equation}\label{Pm}
P_\mu = \int d^3\nad{k}\, (2\omega_{\vec{k}})^{-1} k_\mu \left(a^\dagger(k) 
a(k) + b^\dagger(k) b(k)\right).
\end{equation}
Thus we have defined a consistent Poincar\'e covariant free field theory for a 
fermionic tachyon with helicity $-1/2$; the Weyl's theory 
for a left spinor is obtained as the $\kappa \to 0$ limit. For more 
details of this construction see \cite{Rem96}.

\section{Dynamics}\label{sec:4}
To formulate a dynamical model with tachyonic neutrino we can use, on the 
tree level, the phenomenological form of the weak interaction Lagrangian, 
namely the current-current Fermi Lagrangian. The only condition we take into 
account is that the massless limit $\kappa\to 0$ of our model should coincide 
with the standard four-fermion interaction. This leads to three natural 
possibilities for the part of the lepton charged current containing tachyonic 
neutrino:
\begin{center}
\begin{tabular}{l@{---}l}
$\bar{\ell}(x)\gamma^\mu\frac{1}{2}(1-\gamma^5)\nu(x)$ + h.c. & chiral
coupling, \\
$\bar{\ell}(x)\gamma^\mu\nu(x)$ + h.c. & helicity coupling, \\
$\bar{\ell}(x)\gamma^5\gamma^\mu\nu(x)$ + h.c. & $\gamma^5$ coupling,
\end{tabular}
\end{center}
and similarly for the neutral currents.

Recall that $\nu(x)$ satisfy the helicity condition (\ref{lambda}) with 
$\lambda = 1/2$ so in the $\kappa \to 0$ limit $\nu \to \nu_L$.

However, if we consider higher order processes (like $\nu \to \nu \gamma$) it 
is necessary to have a more sophisticated model like the Salam--Weinberg one. 
Let us notice firstly, that the Lagrangian for the massless Dirac field, 
under assumption of CP conservation, can be written in two different forms 
$\bar\psi i \gamma \partial \psi$ or $\bar\psi \gamma^5 i \gamma \partial 
\psi$. Both forms lead to the same massless Dirac equation $i \gamma 
\partial \psi = 0$ because of invertibility of $\gamma^5$ matrix. Therefore 
generation of a mass {\it via\/} Yukawa coupling with Goldstone fields can 
lead to massive fermions or fermionic tachyons, because of creation in the 
Lagrangian of Dirac $\bar{\psi} \left(i \gamma \partial - m\right) \psi$ or 
$\bar{\psi} \left(\gamma^5 i \gamma \partial - m\right) \psi$ term 
respectively. Thus, because we have decided to treat neutrinos as tachyons, 
we start with the following free massless Lagrangian for a fixed lepton 
generation $(\nu,\ell)$
\begin{equation}
{\cal L}_0 = \bar{\nu} \gamma^5 i \gamma \partial \nu + \bar{\ell} i \gamma 
\partial \ell\,.
\end{equation}

After splitting $\nu$ and $\ell$ into left-handed and right-handed pairs 
$\nu_{L,R}$, $\ell_{L,R}$, it is easy to see that the (compact) invariance 
group of the Lagrangian ${\cal L}_0$ is exactly the $\bigl(SU(2) \times 
U(1)\bigr)_L \times U(1)_R$ group! Therefore, taking into account the 
electric charges of $\nu$ and $\ell$, we conclude that the symmetry group of 
${\cal L}_0$ is $U(1)_{\mbox{\scriptsize lepton number}} \times \bigl(U(1)_Y 
\times SU(2)_I\bigr)$ and the left-handed fields $\nu_L$ and $\ell_L$ must 
form the weak dublet while the right-handed $\nu_R$ and $\ell_R$ are singlets 
of the weak group. {\em Therefore ${\cal L}_0$ leads uniquely to the 
standard Salam--Weinberg choice of the weak symmetry group and their 
realization on the lepton fields\/} (recall also that in the $\kappa \to 0$ 
limit $\nu(\kappa) \to \nu(0) = \nu_L(0)$, i.e. $\nu_R(\kappa) \to \nu_R(0) = 
0$). Now, under the standard choice of the Goldstone fields (as the $U(1)_Y 
\times SU(2)_I$ dublet) as well as the Higgs potential and Yukawa coupling 
(for neutrinos like for neutral quarks) and after gauging the weak group we 
obtain the final Lagrangian
\begin{eqnarray}
\lefteqn{{\cal L} = {\cal L}_{\mbox{\scriptsize fermions + int.}}}\nonumber\\
&&+ \left({\cal L}_{\mbox{\scriptsize gauge bosons}} + 
{\cal L}_{\mbox{\scriptsize Higgs + int.}} + \ldots\right)\,.
\label{Lag}
\end{eqnarray}
The part of ${\cal L}$ in the parentheses has the standard form (we choose 
convention as in \cite{Bai94}) while the ${\cal L}_{\mbox{\scriptsize 
fermions + int.}}$ reads
\begin{eqnarray}
\lefteqn{{\cal L}_{\mbox{\scriptsize fermions + int.}}=\bar{\nu} \gamma^5 i 
\gamma \partial \nu - \kappa \bar{\nu} \nu}\nonumber\\
&& + \bar{\ell} i \gamma \partial 
\ell - m \bar{\ell} \ell + e A_{\mu} \bar{\ell} \gamma^\mu \ell \nonumber\\
&&+ \frac{g}{2 \cos{\theta_W}} Z_\mu \bar{\ell} \gamma^\mu \left(g_V - g_A 
\gamma^5\right) \ell \nonumber\\
&&- \frac{g}{2 \cos{\theta_W}} Z_\mu \bar{\nu} \gamma^\mu 
\frac{1 - \gamma^5}{2} \nu \nonumber\\
&&- \frac{g}{\sqrt{2}} W^{+}_{\mu} \bar{\nu} \gamma^\mu \frac{1 - 
\gamma^5}{2} \ell - \frac{g}{\sqrt{2}} W^{-}_{\mu} \bar{\ell} \gamma^\mu 
\frac{1 - \gamma^5}{2} \nu \nonumber\\
&&- \frac{g}{2 \sqrt2 m_W} \bar{\nu} \left((m - \kappa) I + (m + \kappa) 
\gamma^5\right) \ell G^{+} \nonumber\\
&&- \frac{g}{2 \sqrt2 m_W} \bar{\ell} \left((m - \kappa) I - (m + \kappa) 
\gamma^5\right) \nu G^{-} \nonumber\\
&&+ \frac{i g}{2 m_W} \left(\kappa \bar{\nu} \gamma^5 \nu - m \bar{\ell} 
\gamma^5 \ell\right) G^0 \nonumber\\
&&- \frac{g}{2 m_W} \left(\kappa \bar{\nu} \nu + m 
\bar{\ell} \ell\right) H\,.
\label{L:f}
\end{eqnarray}
Here $g_V = 1/2 - 2 \sin^2{\theta_W}$, $g_A = 1/2$, $g$---weak coupling 
constant, $\theta_W$---Weinberg angle, $G^\pm$ and $G^0$---Goldstone bosons 
and $H$---Higgs field. The corresponding Feynman rules are listed in the 
Appendix~\ref{app:A}.

The following remarks are in order. First of all, the derivation of the 
Lagrangian (\ref{Lag}) with ${\cal L}_{\mbox{\scriptsize fermions + int.}}$ 
given by (\ref{L:f}) is rather heuristic one; the point is that the right 
component of $\nu$, $\nu_R$ goes to zero with $\kappa \to 0$. This means that 
the mass generation mechanism for tachyons is not well defined by Yukawa 
coupling if we start with the massless Lagrangian. Intuitively, it seems that 
the mechanism of mass generation for the tachyonic neutrinos should be caused 
rather by the interaction with gravitational field. This is related to our 
natural assumption about coincidence of the tachyon preferred frame with the 
comoving frame. The next problem is the helicity condition (\ref{lambda}) 
which is necessary for consistency of the free tachyon theory. Of course it 
can be gauged by introduction of the covariant derivatives and causes 
additional relations between multipoint Green functions of the theory. In 
such a case perturbation consistency and renormalizability of this model 
should be proved. We shift these unanswered questions to further 
investigations and here we will treat the Lagrangian (\ref{Lag}) as an 
effective reasonable approximation of a more realistic theory. Because in the
following we do not analyze processes with intrinsic neutrino lines, this 
approximation seems to be correct.

Notice that the Feynman rules derived from (\ref{Lag})--(\ref{L:f}) 
(Appendix~\ref{app:A}), lead to the effective {\em chiral coupling\/} in the 
Fermi Lagrangian. Therefore, in the following all the calculations on the 
tree level will be done for this coupling. However, we will comment results 
obtained with help of the helicity and $\gamma^5$ couplings too.

Tachyonic neutrino is in general unstable even if the leptonic flavors are 
conserved; namely decays of the form {\it tachyon} $\to$ {\it tachyon} + {\it 
sth.} are kinematically admissible. If we take into account the experimental 
upper bounds on the neutrinos masses \cite{Bar96} we can select two dominant 
processes of this type, namely: three body decay  $\nu_\ell \to \nu_\ell + 
\nu_{\ell'} + \bar{\nu}_{\ell'}$ and radiative decay  $\nu_\ell \to \nu_\ell 
+ \gamma$.

\section{Neutrino decay $\nu_\ell \to \nu_\ell \nu_{\ell'} 
\bar{\nu}_{\ell'}$}\label{sec:5}
\begin{figure}
\psfig{figure=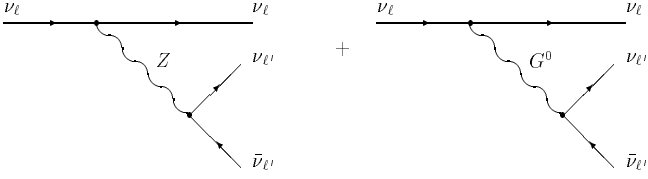,width=8.6cm}
\caption[]{Neutrino three-body decay $\nu_\ell \to \nu_\ell \nu_{\ell'} 
\bar\nu_{\ell'}$}\label{graph:1}
\end{figure}

\begin{figure}
\psfig{figure=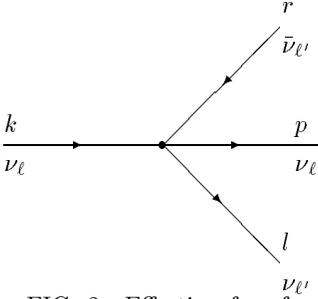}
\caption[]{Effective four-fermion vertex in neutrino three\--body decay
$\nu_\ell 
\to \nu_\ell \nu_{\ell'} \bar\nu_{\ell'}$}\label{graph:2}
\end{figure}

This process can be analyzed on the tree level. The corresponding amplitude 
is the sum of the graphs given in Fig.~\ref{graph:1}
and leads to the effective  four-fermion vertex in Fig.~\ref{graph:2}.
Consequently
\begin{equation}
M = \sqrt{2} G_F \bar{v}_\ell(p) \Gamma^{\mu} {v}_\ell(k) \bar{v}_{\ell'}(l) 
\Gamma_{\mu} {w}_{\ell'}(r)\,,\label{*}
\end{equation}
where the momenta $k$, $p$, $l$ and $r$ correspond to the incoming neutrino 
$\nu_\ell$ and to the outgoing neutrinos $\nu_\ell$, $\nu_{\ell'}$ and 
$\bar{\nu}_{\ell'}$ respectively. The masses of $\nu$ and $\nu_{\ell'}$ are 
denoted by $\kappa$ and $\mu$ respectively. In our case the Feynman rules 
(Appendix~\ref{app:A}) lead to the chiral coupling $\Gamma^{\mu} = (1 + 
\gamma^5) \gamma^{\mu}/2$. Thus, by means of the polarization relations 
(\ref{w-w})--(\ref{v-v}), we obtain in the preferred frame
\begin{eqnarray}
\lefteqn{|M|^2 = \frac{2 G_F^2}{|\vec{k}| |\vec{p}| |\vec{l}| |\vec{r}|}
\bigl(|\vec{k}| + k^0\bigr) 
\bigl(|\vec{p}| + p^0\bigr) 
\bigl(|\vec{l}| + l^0\bigr)}
\nonumber\\
&&\times 
\bigl(|\vec{r}| + r^0\bigr)\left(|\vec{p}| |\vec{l}| - \vec{p} \cdot
\vec{l}\right)
\left(|\vec{k}| |\vec{r}| - \vec{k} \cdot \vec{r}\right)
\label{ampl:ch}
\end{eqnarray}
where from dispersion relations $|\vec{k}| = \sqrt{\kappa^2 + (k^0)^2}$, 
$|\vec{p}| = \sqrt{\kappa^2 + (p^0)^2}$,
$|\vec{l}| = \sqrt{\mu^2 + (l^0)^2}$,  $|\vec{r}| = \sqrt{\mu^2 + (r^0)^2}$.

\begin{figure}
\psfig{figure=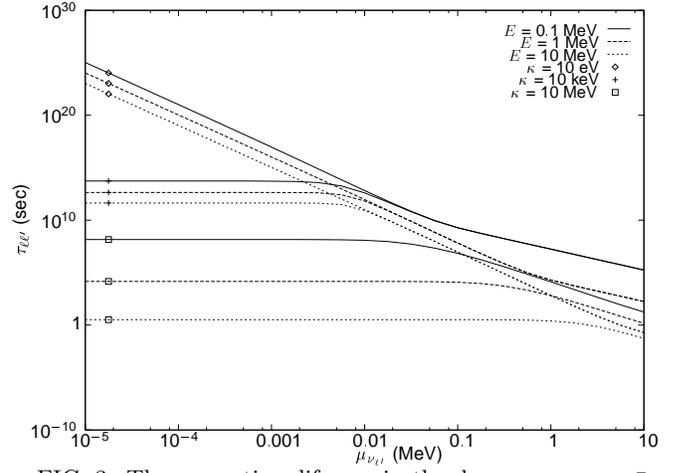,width=8.6cm}
\caption[]{The mean time life $\tau_{\ell \ell'}$ in the decay $\nu_\ell \to 
\nu_\ell \nu_{\ell'} \bar{\nu}_{\ell'}$ as the function of the pair neutrino 
($\nu_{\ell'}$) mass $\mu$ for three different values of the mass $\kappa$ of 
the decaying neutrino $\nu_\ell$. The plot is given for the three values of 
the total energy $E = k^0$ in the preferred frame}\label{fig:times}
\end{figure}

\begin{figure*}
\begin{tabular}{c@{\hspace{.5em}}c}
\psfig{figure=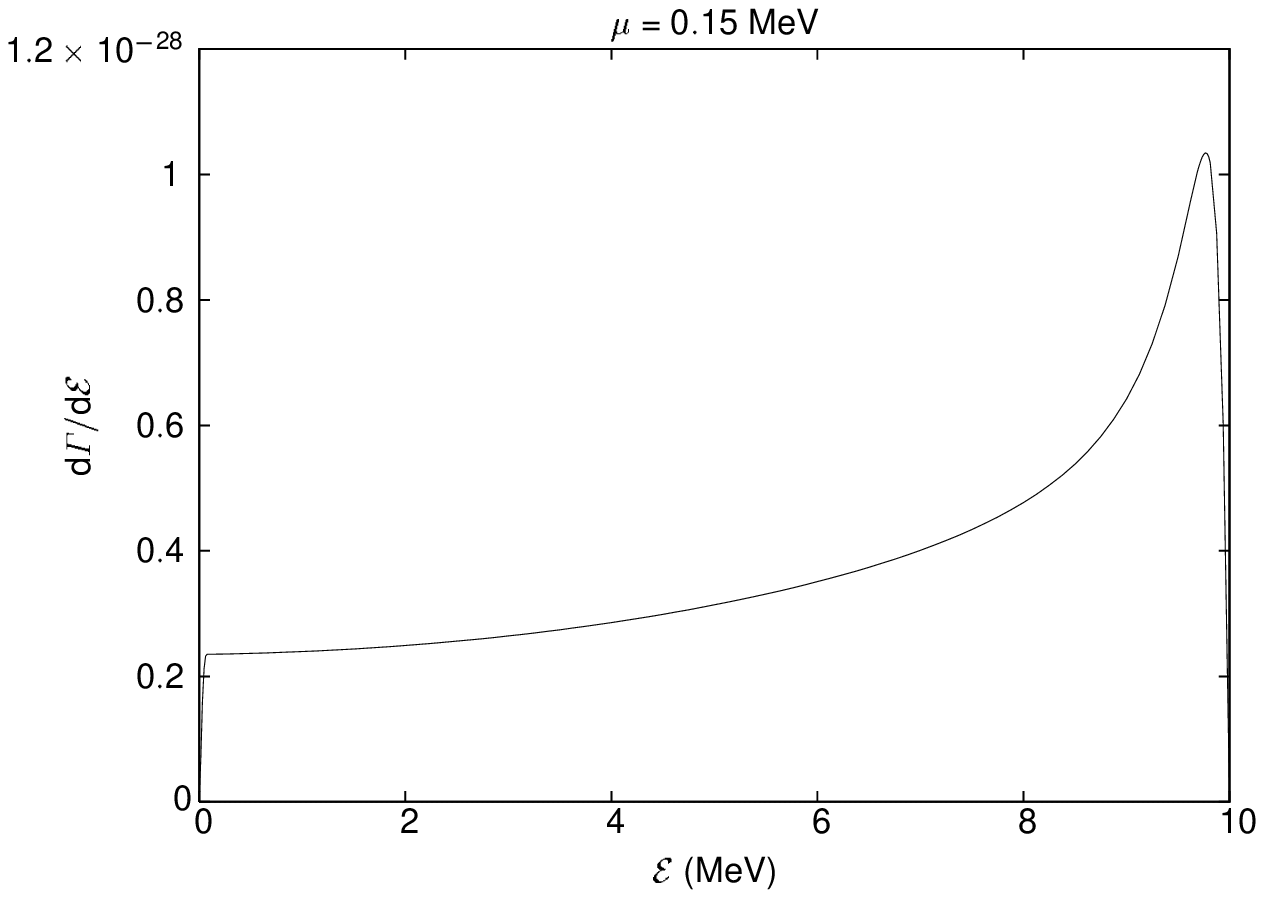,width=8.6cm} &
\psfig{figure=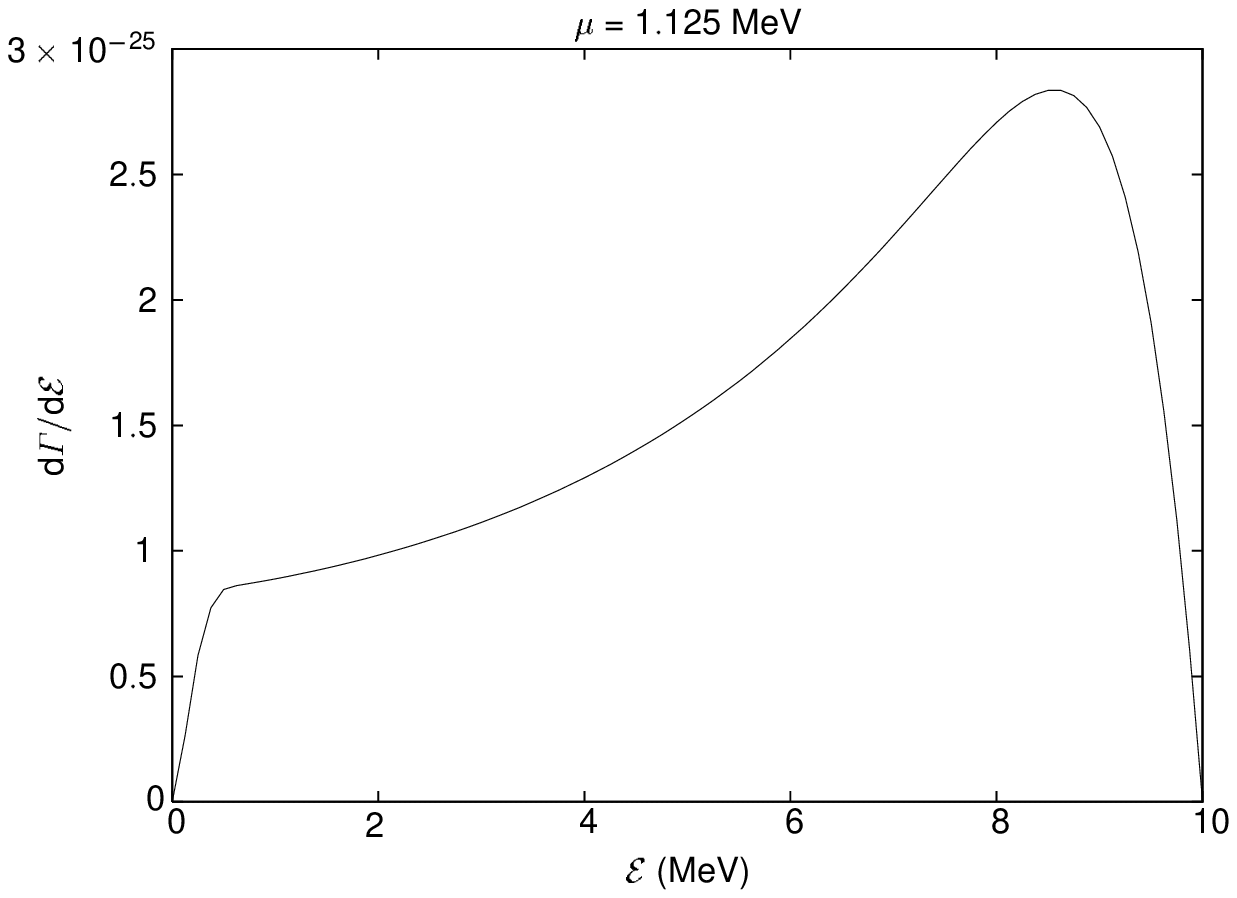,width=8.6cm}
\end{tabular}
\caption[]{Differential decay rate $d\Gamma/d{\cal E}$ for the three-body 
decay $\nu_e \to \nu_e \nu_{\ell'} \bar\nu_{\ell'}$ for $\nu_e$ mass $\kappa 
= 10$~eV as the function of the 
outgoing neutrino $\nu_e$ energy ${\cal E} = p^0$ for $\ell'=\mu$~(on the
left), $\tau$~(on the right), with $\mu_\mu = 0.15$~MeV, $\mu_\tau = 1.125$~MeV 
respectively; the incoming neutrino energy  $k^0 = 10$~MeV}\label{fig:gamma}
\end{figure*}

The general forms of the square of the amplitude for all couplings 
are given in the Appendix~\ref{app:B}. The qualitative behavior of the decay 
rate
\begin{eqnarray}
\lefteqn{\Gamma_{\ell\ell'} = \left((2\pi)^5 2k^0\right)^{-1} 
\int d^4p\, d^4l\, d^4r\, \left|M\right|^2 
\theta(p^0) \theta(l^0) \theta(r^0)} \nonumber\\ 
&\times& \delta(p^2 + \kappa^2) \delta(l^2 + \mu^2) \delta(r^2 + \mu^2) 
\delta^4(k - p - l - r) \label{**}
\end{eqnarray}
is rather similar for all three couplings. Here we
present in Fig.~\ref{fig:times} 
the numerical calculations for the mean life time
in the case of chirality coupling (according to our gauge model) done in
the preferred frame ($u = (1,\vec{0})$). For the helicity coupling the
corresponding results were given in \cite{Cib96a}.
The mean life time of the tachyonic neutrino in this process is
determined by its energy $E = k^0$, as measured in the preferred frame,
and masses of all three neutrino species. As we see from the
Fig.~\ref{fig:times} the life time $\tau_{\ell\ell'}$ ($\ell \neq \ell'$) 
corresponding to the
partial width $\Gamma_{\ell\ell'}$, is a decreasing function of
the initial energy and mass of the neutrino $\nu_{\ell'}$. For
illustration we present in Fig.~\ref{fig:gamma} 
the energy spectrum of
the final neutrino $\nu_\ell$ (after decay) assuming its mass as $\kappa = 
10$~eV and for the $\nu_{\ell'}$, taking $\mu = 0.15$~MeV and $\mu = 
1.125$~MeV. The case $\ell = \ell'$ differs by
the corresponding combinatorial factor.
It is important to stress that:
\begin{itemize}
\item The decay $\nu_\ell \to \nu_\ell \nu_{\ell'} \bar{\nu}_{\ell'}$ has 
character of an emission by $\nu_\ell$ a neutrino--antineutrino pair 
$\nu_{\ell'} \bar{\nu}_{\ell'}$; as a consequence the resulting neutrino flux 
contains neutrinos and antineutrinos of all possible flavors. Therefore this 
process can simulate the neutrino flavor oscillations. Notice that for 
massive and massless neutrinos such a decay is kinematically forbidden.
\item The above process is repeated and forms a cascade. Therefore, as was 
mentioned in the Introduction, it demands more sophisticated treatment as a 
nontrivial stochastic process. However after the neutrino energy degradation 
it is slowing down with subsequent decays; notice that, for low energies in 
the preferred frame, neutrinos are almost stable (Fig.~\ref{fig:times}).
\end{itemize}

\section{Three body decay and the solar neutrino problem}\label{sec:5a}
As is well known, predictions of the Standard Solar Model (SSM) are in a 
strong disagreement with all operating solar neutrino experiments 
\cite{bahcall94,bahcall96,bahcall97a}.

Firstly, the chlorine radiochemical experiment (threshold $\sim 0.81$~MeV) 
gives measured event rate which is a factor $3.6$ less than the theoretically 
predicted by using SSM with helium diffusion\footnote{SSM with helium 
diffusion are mostly favored by the helioseismological experiments 
\cite{bahcall97}.}. The neutrino flux detected in the chlorine experiment is 
from the high energy $^8$B neutrinos with a significant contribution of 
$^7$Be neutrinos; contributions of pep and CNO neutrinos are not so 
significant, \cite{bahcall95}. Secondly, the water experiment (Kamiokande; 
apparatus threshold $\sim 7$~MeV) detects $^8$B neutrinos and gives about 
$0.4$ part of the predicted $^8$B neutrino flux \cite{bahcall97a}. Finally, the 
gallium experiments (GALLEX, SAGE) detects mostly basic $pp$ neutrino flux as 
well as $pep$ neutrinos, but also $^8$B, $^7$Be and CNO fluxes contribute to 
the measured event rate \cite{bahcall95}. Observed event rate in this 
experiment is about $0.52$ of the SSM prediction \cite{bahcall97a}.

Let us check applicability of the tachyonic neutrino three-body decay to the 
solar neutrino problem. Although vacuum oscillation or MSW mechanism are 
possible in this case too, we assume here that neutrino deficit is caused by 
the neutrino decay only. As we will see, this assumption allows us to 
estimate the taonic neutrino mass (if it has tachyonic character) as less 
than $1.125$~MeV. In the following we use the neutrino fluxes as predicted in 
papers by Bahcall and Pinsonneault \cite{bahcall95}.

\begin{figure}
\psfig{figure=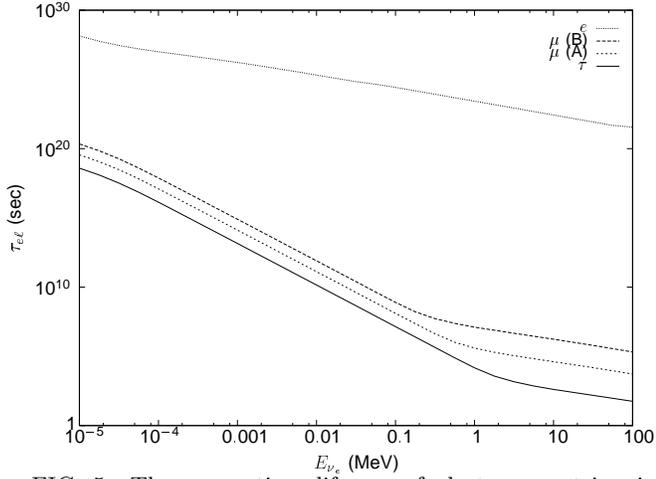,width=8.6cm}
\caption{The mean time life $\tau_{e\ell}$ of electron neutrino in three-body
decay into electron neutrino and neutrino--anti\-neutrino pair of type $\ell$
for
$\ell = e$ (mass $\kappa = \mu_e =\allowbreak 10$~eV), $\ell = \mu$ (A) and (B)
(masses
$\mu_\mu = 0.37$~MeV and $\mu_\mu = 0.15$~MeV corresponding to the Jackelmann A
and B solutions respectively \protect\cite{assamagan96}), and for $\ell = \tau$
(mass $\mu_\tau = 1.125$~MeV)}
\label{fig:tau_en}
\end{figure}

As it is evident from the Fig.~\ref{fig:times}, in the electron neutrino 
($\kappa_e \sim 10$~eV) decay, for energies $1$--$10$~MeV and time 
$\mbox{Sun--Earth distance}/c \sim 500$~sec,
the mass of the resulting $\nu_{\ell'}$ should be of the order $1$~MeV. This 
excludes decays $\nu_e \to \nu_e \nu_e \bar{\nu}_e$ and $\nu_e \to \nu_e 
\nu_\mu \bar{\nu}_\mu$ as possible sources of the investigated effects (recall 
the Jackelmann A and B solutions \cite{assamagan96} for the muon neutrino mass 
$\sim 0.37$~MeV and $\sim 0.15$~MeV respectively). 
We show the dependence of the mean life time of $\nu_e$ on energy for these 
cases on Fig.~\ref{fig:tau_en}.
Therefore in the following we will consider the contribution of the $\nu_e 
\to \nu_e \nu_\tau \bar{\nu}_\tau$ decay only. Our aim is to analyze the 
cascade process
\begin{displaymath}
\begin{array}{ccccc}
\nu_e & \to & \nu_e &\nu_\tau & \bar{\nu}_\tau \\
&&\downarrow&&\\
&& \nu_e &\nu_\tau & \bar{\nu}_\tau \\
&&\downarrow&&\\
&& \nu_e &\nu_\tau & \bar{\nu}_\tau \\
&&\downarrow&&\\
&&\dots&&
\end{array}
\end{displaymath}
taking into account that probabilities of the possible subsequent decays 
$\nu_\tau \to \nu_\tau \nu_e \bar{\nu}_e$ and $\bar{\nu}_\tau \to 
\bar{\nu}_\tau \nu_e \bar{\nu}_e$ are not significant in the investigated 
energy regime.

This greatly simplifies treatment of this stochastic process leading to the 
following evolution equation for the electron neutrino flux:
\begin{equation}\label{dif:flux}
\hbar \frac{d\Phi_t(E)}{d t} = -\Gamma_E \Phi_t(E) + \int_E^\infty d E'\, 
\frac{d\Gamma_{E'}}{d E} \Phi_t(E')\,.
\end{equation}
Hereafter $\Phi_t(E)$ denotes the electron neutrino flux of energy $E$ after 
time $t$, $\Gamma_E$ is the total decay rate for the initial neutrino of 
energy $E$ and $d\Gamma_{E'}/d E$ is the differential decay rate of the 
decaying neutrino $\nu_e$ with energy $E'$ into outgoing electron neutrino 
with energy $E$. The equation (\ref{dif:flux}) has the solution as a formal 
power series
\begin{equation}\label{int:flux}
\Phi_t = e^{t F/\hbar} \Phi_0
\end{equation}
with $F(E,E') = -\Gamma_E \delta(E - E') + \theta(E' - E)\, d\Gamma_{E'}/d
E$.

\begin{figure}
\psfig{figure=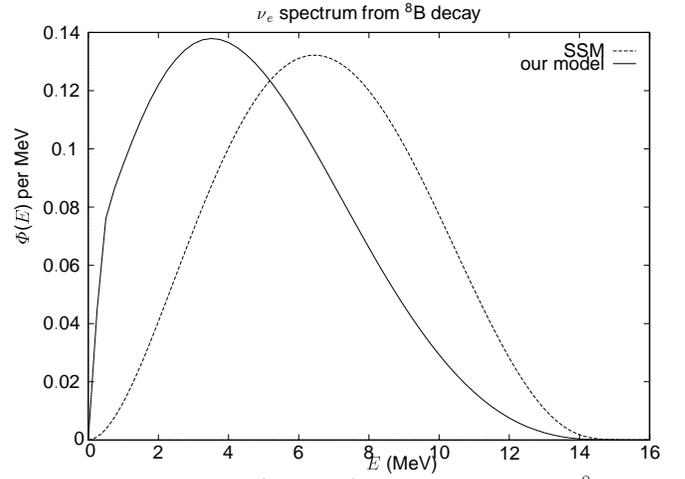,width=8.6cm}
\caption[]{Normalized (to unity) energy spectrum of $^8$B neutrino flux 
registered on the Earth from 
SSM predictions and our model; the masses are assumed to be $\mu_e = 10$~eV,
$\mu_\tau = 1.125$~MeV}\label{fig:8B}
\end{figure}

\begin{figure}
\psfig{figure=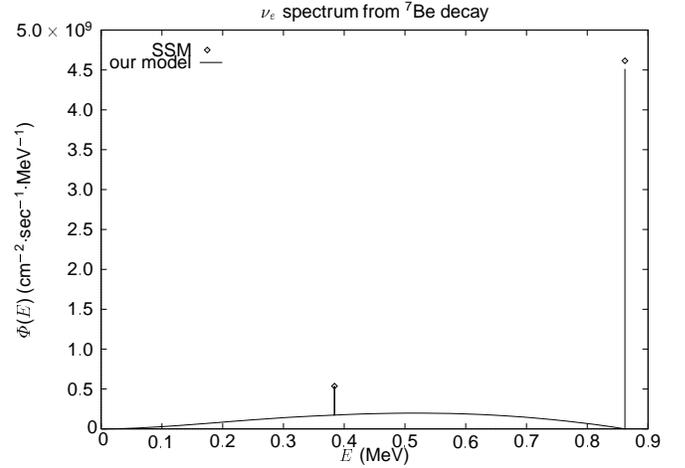,width=8.6cm}
\caption[]{Energy spectrum of $^7$Be neutrino flux from 
SSM predictions and our model; the masses are assumed to be $\mu_e = 10$~eV,
$\mu_\tau = 1.125$~MeV}\label{fig:7Be}
\end{figure}

The numerical estimation of the spectrum of decaying neutrinos with respect to 
the heaviest neutrino mass ($\nu_\tau$) gives the best fit\footnote{For the 
helicity coupling we obtain for $\mu_\tau$ the value $0.95$~MeV.}
for $\mu_\tau = 
1.125$~MeV. In this case the neutrino spectra 
\cite{bahcall95,bahcall86,bahcall88,bahcall92} lead to the results showed 
in the Table~\ref{tab:fl}.
\begin{table*}
\caption[]{Solar neutrino fluxes---comparison of our model and SSM}
\label{tab:fl}
\begin{tabular}{lccccccc}
\noalign{\smallskip}
         & $^2$H 
         & \multicolumn{3}{c}{$^{37}$Cl} 
         & \multicolumn{3}{c}{$^{71}$Ga}\\
         & our model/SSM & our model & SSM & our model/SSM
         & our model& SSM & our model/SSM \\
         &  \%      & (SNU) & (SNU)     & \%          & (SNU) & (SNU)     & \%
         \\
\noalign{\smallskip}\hline\noalign{\smallskip}
$pp$     &          &       &           &             & $69.30$&$69.69$   &
$99$ \\
$pep$    &          & $0.19$&$0.21$     & $93$        & $2.71$&$2.86$     &
$95$ \\
$^7$Be   &          & $1.38$&$1.41$     & $98$        & $37.37$&$37.82$   &
$99$ \\
$^8$B    & $41$     & $2.83$&$7.01$     & $40$        & $7.35$&$16.16$    &
$45$ \\
$^{13}$N &          & $0.10$&$0.11$     & $94$        & $3.77$&$3.88$     &
$97$ \\
$^{15}$O &          & $0.34$&$0.36$     & $95$        & $6.13$&$6.33$     &
$97$ \\
$^{17}$F &          & $0.0041$&$0.0043$ & $95$        & $0.073$&$0.076$   &
$97$ \\
total    & $41$     & $4.84$&$9.10$     & $53$        & $126.70$&$136.82$ &
$93$ \\
our model $-$ 
$^7$Be  & $41$     & $3.46$&$9.10$     & $38$        & $89.33$&$136.82$  & $65$
\\
\noalign{\smallskip}
\end{tabular}
\end{table*}
It is evident that, under assumption of the correctness of the SSM predictions, 
the three-body neutrino decay  does not solve the problem to the end. 
However we can see that the problems mentioned above 
can be reduced to the overflow of $^7$Be neutrinos only. Indeed, if we set 
$^7$Be flux zero, we obtain approximately $\Phi_{pp} \sim \Phi_{pp}^{\rm 
SSM}$ and $\Phi_{^8{\rm B}} \sim 0.4 \Phi_{^8{\rm B}}^{\rm SSM}$, 
consistently with observed values. We would point out, that in \cite{cumming96}
it was argued that a slow mixing of $^3$He down into the solar core can cause
significant reduction of the $^7$Be flux. On the 
Fig.~\ref{fig:8B} and Fig.~\ref{fig:7Be} we compare calculated by ourselves 
and predicted by SSM $^8$B and $^7$Be neutrino fluxes.
In principle, the change of the shape of $^8$B neutrino flux (if it 
actually changes), can be measured in the SuperKamiokande experiment.

Finally, let us stress that the above results was obtained under the 
assumption of the absence of possible neutrino oscillations But they can be
used to an estimation of the upper bound for the taonic neutrino mass. We
intend to return to the more general situation (with oscillations) in the
forthcoming paper.

\section{Radiative decay of the tachyonic neutrino $\nu_\ell \to \nu_\ell + 
\gamma$}\label{sec:6}
This process can play an important role in cosmology; in particular, as was 
noticed in \cite{fukugita88,fukugita89,maalampi89}, it can affect the cosmic 
background radiation in the submilimeter region, leading to its distortion.

Let us denote the four-momenta of the initial and final neutrinos by $k$ and 
$p$ respectively, their masses by $\kappa$, while photon has the 
four-momentum $q = k - p$. The 
kinematics of this process is very simple; namely $k^2 = -\kappa^2$, $p^2 = 
-\kappa^2$ and for the photon on the mass shell $q^2 = (k - p)^2 = 0$ so $k p 
= -\kappa^2$, $q k = q p = 0$. Let us consider this kinematics in the 
preferred frame $u = (1,\vec{0})$. It is immediate to see that 
$\cos\theta = \vec{k} \cdot \vec{q}/(|\vec{k}| |\vec{q}|) = k^0 \left((k^0)^2 
+ \kappa^2\right)^{-1/2}$. Therefore the outgoing photons 
are emitted on the cone determined by the angle $2\theta$. Notice that for 
$k^0 \gg \kappa$, $\cos{\theta} \approx 1$ so the process is approximately 
collinear in that case.

The amplitude has the form
\begin{equation}
M^{\lambda}(p,k) = \Omega^{\mu}(p,k) e_{\mu}^{\lambda}(q)
\end{equation}
where $e_{\mu}^{\lambda}$ is the polarization vector for the photon with 
helicity $\lambda$. The $\Omega^{\mu}(p,k)$ should satisfy transversality and 
neutrality conditions. The first one follows from the electromagnetic current 
conservation and has the form
\begin{equation}
q_{\mu} \Omega^{\mu}(p,k) = 0  \label{trans}
\end{equation}
while neutrality condition means that the neutrino charge is zero and 
consequently the corresponding matrix element vanishes:
\begin{equation}
\Omega^{\mu}(k,k) = 0.    \label{neutral}
\end{equation}
By means of the equations of motion for the spinors $\bar{v}$ and $v$ 
(\ref{E3}), the most general form of the $\Omega^{\mu}(p,k)$ reads
\begin{eqnarray}
\lefteqn{\Omega^{\mu}(p,k) = \bar{v}(p) \left\{(F_V + F_A\gamma^5) 
\gamma^{\mu}\right.} \nonumber\\
&&+ \left.(G_{1V} + i G_{1A} \gamma^5) k^{\mu} \right. \nonumber\\
&&+ \left.(G_{2V} + i G_{2A} \gamma^5) p^{\mu} \right\} {v}(k)\,,
\label{4}
\end{eqnarray}
where the form factors $F$ and $G$ are in general functions of $q^2$. Notice 
that the form factors $G_{iA}$ control CP non-invariant terms in the 
Eq.~(\ref{4}). Now, taking into account the transversality (\ref{trans}) and 
neutrality (\ref{neutral}) we obtain the following constraints
\begin{equation}
\begin{array}{l}
\displaystyle F_V = 0\,, \\
\displaystyle F_A = \kappa (G_{1V} + G_{2V})\,, \\
\displaystyle (G_{2A} - G_{1A}) q^2 = 0\,, \\
\displaystyle (G_{2V} - G_{1V}) q^2 = 0\,.
\end{array}
\end{equation}
Therefore, because of the analyticity of $G_{iV}$ and $G_{iA}$ in the point 
$q^2 = 0$, we obtain
\begin{equation}
\Omega^{\mu}(p,k) = -\frac{1}{2} \bar{v}(p) \left[\gamma^{\mu}, 
\gamma^{\nu}\right] q_{\nu} (G_V + i G_A \gamma^5) v(k)\,,
\end{equation}
where we denote $G_V \equiv G_{1V} = G_{2V}$, $G_A \equiv G_{1A} = G_{2A}$ 
and the equations of motion (\ref{E3}) was used again. Therefore the square 
of the amplitude $M^{\lambda}$, after summation over final polarizations of 
the photon (Appendix~\ref{app:D}), takes the form:
\begin{eqnarray}
\lefteqn{\left|M\right|^2 = 4 \kappa^4 \left(\left|G_V\right|^2 + 
\left|G_A\right|^2\right)} \nonumber\\
&\times& \frac{(uq)^2}{\sqrt{(u (k - q))^2 + \kappa^2} \sqrt{(u k)^2 + 
\kappa^2}} 
\label{10}
\end{eqnarray}
where we have used the form of the polarization operator
(\ref{v-v}).

\begin{figure*}
\begin{center}\psfig{figure=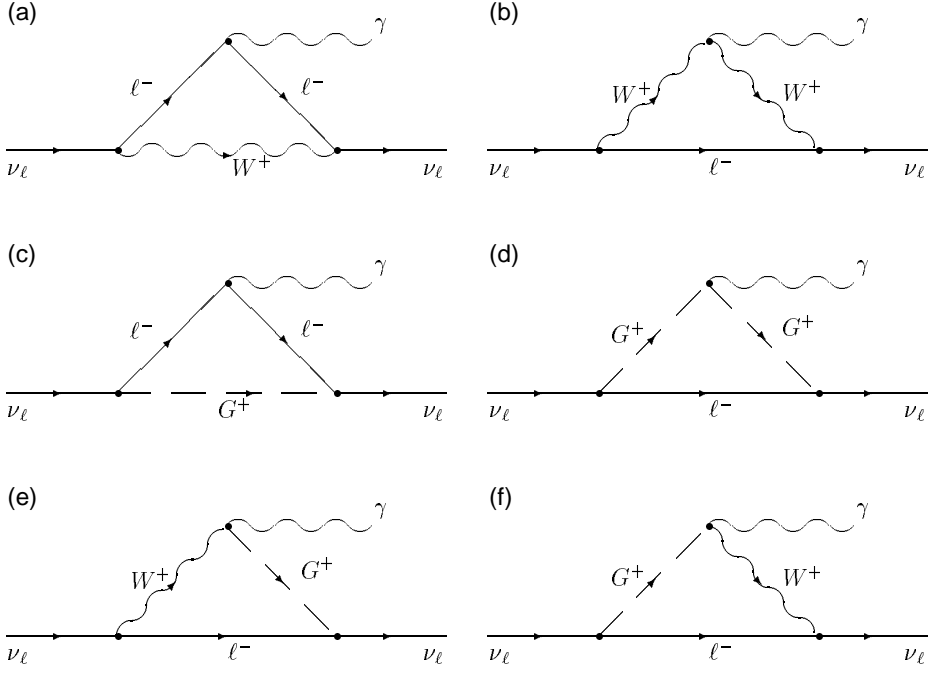,width=12.3cm}\end{center}
\caption[]{Contributions to the process $\nu_\ell \to \nu_\ell + \gamma$ to 
one-loop order}\label{graph:3}
\end{figure*}

In the framework of our $SU(2)\times U(1)$ gauge model, the diagrams 
contributing to the the process $\nu_\ell \to \nu_\ell + \gamma$ to one-loop 
order are shown on Fig.~\ref{graph:3}.

Now, the transversality and neutrality conditions (\ref{trans}), 
(\ref{neutral}) can be treated as renormalization conditions. By means of the 
Feynman rules listed in the Appendix A we can easily calculate the form 
factors $G_A$ and $G_V$:
\begin{equation} \label{GA}
G_A  =  0,
\end{equation}
as we expected because of the CP invariance of this model, while
\begin{equation} \label{GV}
G_V = \frac{e g^2\kappa}{64 \pi^2} \left[I_a + I_b + I_c + I_d + I_e + 
I_f\right]
\end{equation}
where (see Appendix~\ref{app:C}), under the experimental conditions 
for lepton masses $m \ll m_W$ and for corresponding neutrino masses 
$\kappa \ll m_W$,
\begin{eqnarray*}
I_a &=& \frac{4}{3 m_W^2}\,, \\
I_b &=& -\frac{7}{6 m_W^2}\,, \\
I_c &=& \frac{2 m^2}{m_W^4} \ln\left(\frac{m_W}{m}\right)^2\,, \\
I_d &=& \frac{5 m^2 + \kappa^2}{6 m_W^4}\,, \\
I_e + I_f &=&  \frac{1}{2 m_W^2}\,.
\end{eqnarray*}
Thus in this approximation $I_c + I_d$ can be 
omitted and the final form of $G_V$ reads
\begin{equation}
G_V = \frac{e g^2 \kappa}{96 \pi^2 m_{W}^{2}} = \frac{\kappa G_F}{6 \pi} 
\sqrt{\frac{\alpha}{2 \pi}}\,.   \label{Lk}
\end{equation}
Consequently
\begin{equation}
\Omega^{\mu} = -\frac{\kappa G_F}{12 \pi}\sqrt{\frac{\alpha}{2 \pi}} 
\bar{v}(p) \left[\gamma^{\mu}, \gamma^{\nu}\right] q_{\nu} v(k)
\label{amplitude}
\end{equation}
and
\begin{equation}
\left|M\right|^2 = \frac{\alpha \kappa^6 G_{F}^{2}}{18 \pi^3} 
\frac{(u q)^2}{\sqrt{(u k)^2 + \kappa^2}\sqrt{(u (k - q))^2 + \kappa^2}}\,.
\label{amplitudesqr}
\end{equation}
Notice, that (\ref{amplitude}) implies that the magnetic moment $\mu_{\nu}$ 
of the neutrino $\nu$ is given by
\begin{equation}
\mu_{\nu} = \frac{\kappa G_F}{3 \pi}\sqrt{\frac{\alpha}{2 \pi}} = 
\frac{G_F m_e \kappa}{3 \sqrt{2} \pi^2} \mu_B\,,
\label{moment}
\end{equation}
where $\mu_B$ is the Bohr magneton and $m_e$ the electron mass. (N.B.\ the 
value of $\mu_\nu$ is of the same form as for the massive neutrino; the 
difference lies in the numerical factor only \cite{Moh91}). To calculate 
the decay rate we integrate $\left|M\right|^2$ over the phase space: 
\begin{eqnarray}
\lefteqn{\Gamma = (8 \pi^2 k^0)^{-1} \int d^4q \,d^4p \left|M\right|^2 
\theta(q^0) \theta(p^0)} \nonumber\\
&\times& \delta(q^2) \delta(p^2 + \kappa^2) \delta(k - p - q)\,.
\label{25}
\end{eqnarray}
We shall do it in the preferred frame, that is we put $u = (1,\vec{0})$.
Inserting $\left|M\right|^2$ given by (\ref{amplitudesqr})
we obtain the energy spectrum of emitted photons as:
\begin{equation}
\frac{d\Gamma}{d q^0} = 
\frac{\alpha G_{F}^{2} \kappa^6 (q^0)^2 \theta(k^0 - q^0) \theta(q^0)}
{288 \pi^4 k^0 \left((k^0)^2 + \kappa^2\right) \sqrt{(k^0 - q^0)^2 + 
\kappa^2}}\,.
\end{equation}
Finally, the total decay rate reads:
\begin{eqnarray}
\lefteqn{\Gamma = \frac{\alpha \kappa^6 G_F^2}{576 \pi^4 k^0 \left((k^0)^2 + 
\kappa^2\right)} \left(4 \kappa k^0 - 3 k^0 \sqrt{(k^0)^2 + \kappa^2} 
\vphantom{\ln\frac{\sqrt{(k^0)^2 + \kappa^2} + k^0}{\kappa}} 
\right.} 
\nonumber\\
&+& \left.\left(2 (k^0)^2 - \kappa^2\right) 
\ln\frac{\sqrt{(k^0)^2 + \kappa^2} + k^0}{\kappa}\right)\,.
\label{przekroj}
\end{eqnarray}

Note that the experimental upper bounds on the neutrino masses 
applied to the magnetic moment formula (\ref{moment}) give results which are 
in agreement with the experimental data \cite{Bar96} (cf.~Table~\ref{tab:0}).
\begin{table}
\caption[]{Upper bounds on the neutrino magnetic moment}\label{tab:0}
\begin{tabular}{ccc}
\noalign{\smallskip}
& from (\ref{moment}) & from \cite{Bar96} \\
\noalign{\smallskip}\hline
\noalign{\smallskip}
$\mu_e$     & $<1.4\times10^{-18} \mu_B$ & $<1.8\times10^{-10} \mu_B$ \\
$\mu_\mu$   & $<2.4\times10^{-14} \mu_B$ & $<7.4\times10^{-10} \mu_B$ \\
$\mu_\tau$  & $<3.4\times10^{-12} \mu_B$ & $<5.4\times10^{-7}  \mu_B$ \\
\noalign{\smallskip}
\end{tabular}
\end{table}

\begin{figure}
\psfig{figure=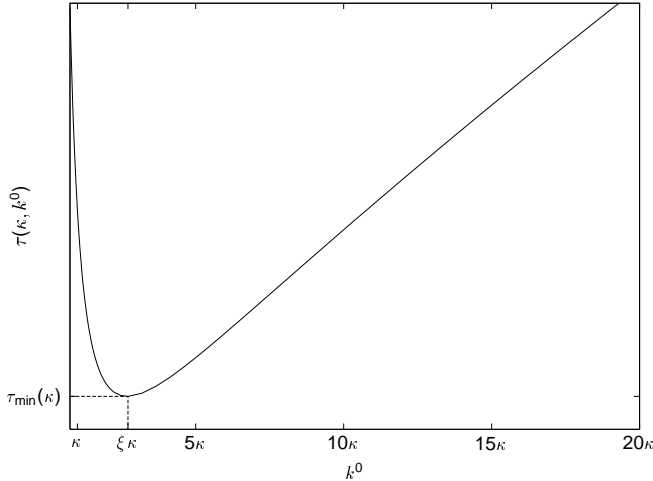,width=8.6cm}
\caption[]{Mean life time $\tau$ in the radiative decay $\nu \to \nu \gamma$ 
as the function of the neutrino mass $\kappa$ and the initial energy $k^0$. 
Here $\xi \approx 2.6899$}\label{fig:tau}
\end{figure}

\begin{figure}
\psfig{figure=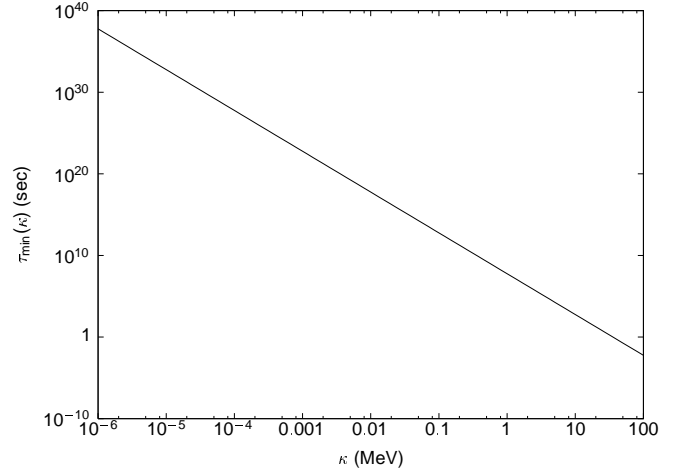,width=8.6cm}
\caption[]{Minimal mean time of life $\tau_{\mathrm{min}}$ for the
radiative decay $\nu \to \nu \gamma$ as the function of the neutrino
mass $\kappa$}\label{fig:tau_min}
\end{figure}

\begin{figure}
\psfig{figure=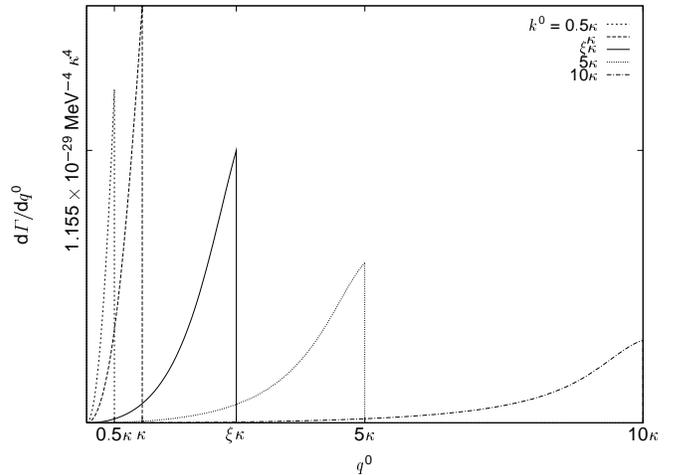,width=8.6cm}
\caption[]{Differential rate $d\Gamma/d q^0$ for the radiative decay $\nu 
\to \nu \gamma$ as the function of the photon energy $q^0$ close to the 
minimum of the mean time of life}\label{fig:5}
\end{figure}

Now, for illustration we present the mean life time $\tau = \hbar/\Gamma$ for 
this process as a function of the energy $k^0$ in Fig.~\ref{fig:tau}. 
From (\ref{przekroj}) it is easy to see that for $k^0 = \xi \kappa$, with 
$\xi \approx 2.6899$ the mean time of life has a minimum. The minimal value 
$\tau_{\rm min}$ as the function of the mass $\kappa$ reads
\begin{equation}
\tau_{\rm min}(\kappa) \simeq \frac{902 \pi^4 \hbar}{\alpha G_F^2} \kappa^{-5}
\end{equation}
and is presented in Fig.~\ref{fig:tau_min}. 
Finally, in Fig.~\ref{fig:5} 
differential rate $d\Gamma/d q^0$ as the function of the photon energy 
$q^0$ and the initial energy $k^0$ is given.

\section{$\beta$-decay with tachyonic antineutrino}\label{sec:7}
\begin{figure}
\psfig{figure=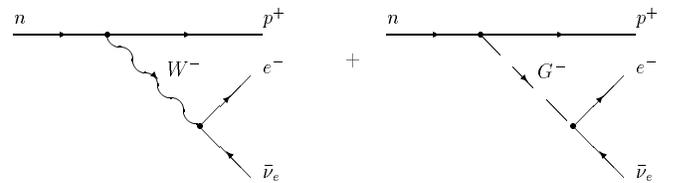,width=8.6cm}
\caption[]{Graphs contributing to the $\beta$-decay}\label{graph:4}
\end{figure}

The tritium $\beta$-decay with tachyonic antineutrino was discussed in the 
papers by Ciborowski and Rembieli\'nski \cite{Cib96a,Cib96b}. Here for a 
completeness we present only a brief view on results obtained in 
\cite{Cib96b}. As in the standard case, the two graphs shown on the 
Fig.~\ref{graph:4} 
contribute to this process on the tree level, where 
dominant is the first one. Effectively, for energies much less than $m_W$, it 
reduces to the four-fermion interaction with the amplitude square given by 
chiral coupling
\begin{eqnarray}
\lefteqn{\left|M_{\rm ch}\right|^2 = 2 G_F^2 {\rm Tr} \left[u_e \bar{u}_e 
\gamma^\mu \frac{1 - \gamma^5}{2} w \bar{w} \gamma^\nu \frac{1 - 
\gamma^5}{2}\right]} \nonumber\\
&\times& {\rm Tr} \left[u_p \bar{u}_p \gamma_\mu (1 - g_A \gamma^5) u_n 
\bar{u}_n \gamma_\nu (1 - g_A \gamma^5)\right]
\end{eqnarray}
where the axial coupling $g_A \approx 1.25$; in the case of helicity coupling,
factors $(1 - \gamma^5)/2$ are replaced by $1$. The explicit calculation of the 
electron spectrum for chiral and helicity couplings as well as comparision with
the experimental data are given in \cite{Cib96b}. 
Let us stress only that the mysterious bump near the 
end point is reproduced on the theoretical plot. More exhaustive discussion of
this question, as well as of the second anomaly problem in the tritium
$\beta$-decay can be solved with help of the tachyonic hypothesis
\cite{Cib96b}. 

\section{Conclusions}\label{sec:8}
In this paper we have analyzed the introductory results related to tachyonic 
neutrino hypothesis. We do not pretend to explain all the problems with 
neutrino physics by means of this hypothesis. However, from the behaviour of 
the calculated decay rates for three dominant 
processes: emission of a $\nu \bar\nu$ pair by a neutrino, radiative decay 
$\nu \to \nu \gamma$ and $\beta$-decay, it follows that this possibility 
cannot be ruled out of considerations. Moreover such a hypothesis provides us 
with an additional mechanism simulating flavor oscillations. The calculations 
involving both oscillations and three body decay for solar, atmospheric 
and SN1987A neutrino fluxes are now in progress.

\section*{Acknowledgement}
One of us (J.R.) thanks to Jacek Ciborowski for interesting discussions.

\appendix
\section{Vertices in our gauge model}\label{app:A}
The vertices necessary for calculation of amplitudes for three processes 
considered here in our gauge model are shown on the Table~\ref{tab:1}. 
\begin{table*}
\caption[]{Vertices in our gauge model used in this work}\label{tab:1}
\begin{tabular}{cl@{\hspace{1em}}cl}
\noalign{\smallskip}
\parbox{60pt}{\psfig{figure=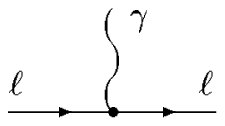}} & 
$\displaystyle i e \gamma^\mu$ &
\parbox{60pt}{\psfig{figure=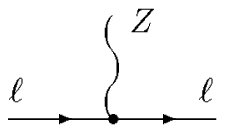}} & 
$\displaystyle\frac{i g \gamma^\mu}{2 \cos\theta_W} 
\left(g_V - g_A\gamma^5\right)$ \\
\noalign{\smallskip}
\parbox{60pt}{\psfig{figure=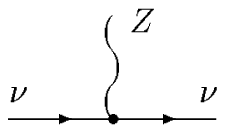}} &
$\displaystyle-\frac{i g \gamma^\mu}{2 \cos\theta_W} 
\frac{1 - \gamma^5}{2}$ &
\parbox{60pt}{\psfig{figure=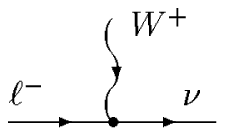}} &
$\displaystyle-\frac{i g \gamma^\mu}{\sqrt{2}} \frac{1 - \gamma^5}{2}$ \\
\noalign{\smallskip}
\parbox{60pt}{\psfig{figure=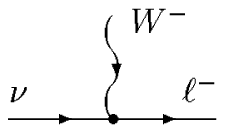}} &
$\displaystyle-\frac{i g \gamma^\mu}{\sqrt{2}} \frac{1 - \gamma^5}{2}$ &
\parbox{60pt}{\psfig{figure=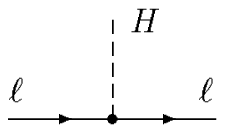}} &
$\displaystyle-\frac{i g m}{2 m_W}$ \\
\noalign{\smallskip}
\parbox{60pt}{\psfig{figure=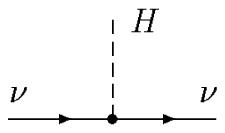}} &
$\displaystyle-\frac{i g \kappa}{2 m_W}$ &
\parbox{60pt}{\psfig{figure=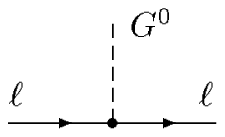}} &
$\displaystyle\frac{g m \gamma^5}{2 m_W}$ \\
\noalign{\smallskip}
\parbox{60pt}{\psfig{figure=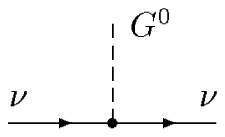}} &
$\displaystyle-\frac{g \kappa \gamma^5}{2 m_W}$ &
\parbox{60pt}{\psfig{figure=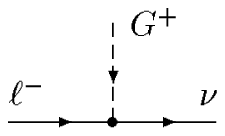}} &
$\displaystyle-\frac{i g}{2 \sqrt2 m_W} \left[(m - \kappa) I 
+ (m + \kappa) \gamma^5\right]$ \\
\noalign{\smallskip}
\parbox{60pt}{\psfig{figure=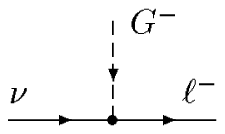}} &
$\displaystyle-\frac{i g}{2 \sqrt2 m_W} \left[(m - \kappa) I 
- (m + \kappa) \gamma^5\right]$ &
\parbox{60pt}{\psfig{figure=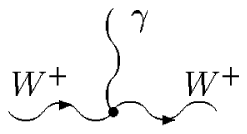}} &
$\displaystyle i e \left[(r - q)_\lambda g_{\mu\nu} 
+ (q - p)_\nu g_{\lambda\mu} + (p - r)_\mu g_{\nu\lambda}\right]$ \\
\noalign{\smallskip}
\parbox{60pt}{\psfig{figure=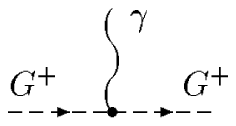}} &
$\displaystyle-i e (p + q)_\mu$ &
\parbox{60pt}{\psfig{figure=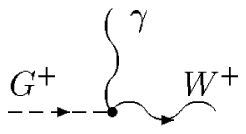}} &
$\displaystyle i e m_W g_{\mu\nu}$ \\
\noalign{\smallskip}
\end{tabular}
\end{table*}
The rest of the Feynman rules in the convention adopted in this paper is 
given e.g.\ in the Ref.~\cite{Bai94}.

\section{Amplitudes for neutrino decay $\nu_\ell \to \nu_\ell 
\nu_{\ell'} \bar\nu_{\ell'}$ for all couplings}\label{app:B}
The square of amplitude for the chiral coupling 
\begin{equation}
\left|M\right|^2 = 2 G_F^2 m^{\mu\nu} n_{\mu\nu} \label{chiral}
\end{equation}
where
\begin{eqnarray*}
\lefteqn{m^{\mu\nu} = \frac{1}{2} 
\left(1 + \frac{(uk)}{\sqrt{(uk)^2 + \kappa^2}}\right) 
\left(1 + \frac{(up)}{\sqrt{(up)^2 + \kappa^2}}\right)} \\
&&\left(k^{\mu} p^{\nu} + k^{\nu} p^{\mu} - (kp) g^{\mu\nu} 
- i \varepsilon^{\mu\nu\alpha\beta} k_{\alpha} p_{\beta}\right) \\
&+&  \frac{\kappa^2}{2 \sqrt{(up)^2 + \kappa^2}} 
\left(1 + \frac{(uk)}{\sqrt{(uk)^2 +\kappa^2}}\right) 
\left(k^{\mu} u^{\nu}\vphantom{\varepsilon^{\mu\nu\alpha\beta}}\right.\\
&&\mbox{}\quad + \left.k^{\nu} u^{\mu} - (uk) g^{\mu\nu}
- i \varepsilon^{\mu\nu\alpha\beta} k_{\alpha} u_{\beta}\right) \\
&+& \frac{\kappa^2}{2 \sqrt{(uk)^2 + \kappa^2}}
\left(1 + \frac{(up)}{\sqrt{(up)^2 + \kappa^2}}\right) 
\left(p^{\mu} u^{\nu}\vphantom{\varepsilon^{\mu\nu\alpha\beta}}\right.\\
&&\mbox{}\quad + \left.p^{\nu} u^{\mu} - (up) g^{\mu\nu}
+ i \varepsilon^{\mu\nu\alpha\beta} p_{\alpha} u_{\beta}\right) \\
&+& \frac{\kappa^4}{2 \sqrt{(uk)^2 + \kappa^2}\sqrt{(up)^2 + \kappa^2}}
\left(2 u^{\mu} u^{\nu} - g^{\mu\nu}\right)
\end{eqnarray*}
and
\begin{eqnarray*}
\lefteqn{n_{\mu\nu} = \frac{1}{2}
\left(1 + \frac{(ul)}{\sqrt{(ul)^2 + \mu^2}}\right)
\left(1 + \frac{(ur)}{\sqrt{(ur)^2 + \mu^2}}\right)} \\
&&\left(l_{\mu} r_{\nu} + l_{\nu} r_{\mu} - (lr) g_{\mu\nu}
+ i \varepsilon_{\mu\nu\alpha\beta} l^{\alpha} r^{\beta}\right) \\
&+& \frac{\mu^2}{2 \sqrt{(ur)^2 + \mu^2}}
\left(1 + \frac{(ul)}{\sqrt{(ul)^2 + \mu^2}}\right) 
\left(l_{\mu} u_{\nu}\vphantom{\varepsilon_{\mu\nu\alpha\beta}}\right.\\
&&\mbox{}\quad + \left.l_{\nu} u_{\mu} - (ul) g_{\mu\nu}
+ i \varepsilon_{\mu\nu\alpha\beta} l^{\alpha} u^{\beta}\right)\\
&+& \frac{\mu^2}{2 \sqrt{(ul)^2 + \mu^2}}
\left(1 + \frac{(ur)}{\sqrt{(ur)^2 + \mu^2}}\right)
\left(r_{\mu} u_{\nu}\vphantom{\varepsilon_{\mu\nu\alpha\beta}}\right.\\
&&\mbox{}\quad + \left.r_{\nu} u_{\mu} - (ur) g_{\mu\nu}
- i \varepsilon_{\mu\nu\alpha\beta} r^{\alpha} u^{\beta}\right)\\
&+& \frac{\mu^4}{2 \sqrt{(ul)^2 + \mu^2}\sqrt{(ur)^2 + \mu^2}}
\left(2u_{\mu} u_{\nu} - g_{\mu\nu}\right)\,.
\end{eqnarray*}

Squares of amplitude for helicity ($s = 0$) and $\gamma^5$ ($s = 
1$) coupling read
\begin{equation}
\left|M\right|^2=2G_{F}^{2}m^{\mu\nu}_{s}n_{s\mu\nu}\,,
\end{equation}
where
\begin{eqnarray*}
\lefteqn{m^{\mu\nu}_{s} = g^{\mu\nu}(u) \left[\left((-1)^s \kappa^2 - k
p\right) 
\right.}\\
&&\mbox{}\quad + \frac{1}{\sqrt{(u k)^2 + \kappa^2} \sqrt{(u p)^2 + \kappa^2}}
\left((-1)^s \kappa^2 \left((k p)\right.\right.\\
&&\mbox{}\qquad \left.- (u k) (u p)\right) - \kappa^4 - \kappa^2 
\left((u p)^2 + (u k)^2 \right) \\
&&\mbox{}\qquad \left.\left.- (u k) (u p) (k p)
\vphantom{(-1)^s}\right)\right]\\
&+& u^{\mu} u^{\nu} \frac{2 \kappa^2 \left(\kappa^2 - (-1)^s (k p)\right)}
{\sqrt{(u k)^2 + \kappa^2} \sqrt{(u p)^2 + \kappa^2}} \\
&+& \left(k^{\mu} p^{\nu} + k^{\nu} p^{\mu}\right) 
\left[\frac{(u k) (u p) - (-1)^s \kappa^2}
{\sqrt{(u k)^2 + \kappa^2}\sqrt{(u p)^2 + \kappa^2}} + 1\right]\\
&+& \left(u^{\mu} k^{\nu} + u^{\nu} k^{\mu}\right)
\frac{\kappa^2 \left((-1)^s (u p) + (u k)\right)}
{\sqrt{(u k)^2 + \kappa^2} \sqrt{(u p)^2 + \kappa^2}}\\
&+& \left(u^{\mu} p^{\nu} + u^{\nu} p^{\mu}\right)
\frac{\kappa^2 \left((-1)^s (u k) + (u p)\right)}
{\sqrt{(u k)^2 + \kappa^2} \sqrt{(u p)^2 + \kappa^2}}\\
&+& i \varepsilon^{\mu\nu}_{~~\alpha\beta}
\left[\frac{\kappa^2\left((-1)^s k^{\alpha}+p^{\alpha}\right)u^{\beta}
-(uk)k^{\alpha}p^{\beta}}{\sqrt{(uk)^2+\kappa^2}}\right.\\
&&\mbox{}\quad\left.-\frac{\kappa^2\left((-1)^s
p^{\alpha}+k^{\alpha}\right)u^{\beta}
+(up)k^{\alpha}p^{\beta}}{\sqrt{(up)^2+\kappa^2}}\right]
\end{eqnarray*}
and
\begin{eqnarray*}
\lefteqn{-n_{s\mu\nu}=g_{\mu\nu}(u)\left[\left((-1)^s\mu^2+lr\right)\right.}\\
&&\mbox{}\quad+\frac{1}{\sqrt{(ul)^2+\mu^2}\sqrt{(ur)^2+\mu^2}}
\left((-1)^s\mu^2\left((lr)\right.\right.\\
&&\mbox{}\qquad\left.-(ul)(ur)\right)+\mu^4+\mu^2
\left((ul)^2+(ur)^2\right)\\
&&\mbox{}\qquad\left.\left.+(ul)(ur)(lr)\vphantom{(-1)^s}\right)\right]\\
&+& u_{\mu}u_{\nu}\frac{2\mu^2\left(-\mu^2-(-1)^s(lr)\right)}
{\sqrt{(ul)^2+\mu^2}\sqrt{(ur)^2+\mu^2}}\\
&+&\left(l_{\mu}r_{\nu}+l_{\nu}r_{\mu}\right)
\left[-\frac{(ul)(ur)-(-1)^s\mu^2}
{\sqrt{(ul)^2+\mu^2}\sqrt{(ur)^2+\mu^2}}-1\right]\\
&+&\left(u_{\mu}l_{\nu}+u_{\nu}l_{\mu}\right)
\frac{\mu^2\left((-1)^s(ur)-(ul)\right)}
{\sqrt{(ul)^2+\mu^2}\sqrt{(ur)^2+\mu^2}}\\
&+&\left(u_{\mu}r_{\nu}+u_{\nu}r_{\mu}\right)
\frac{\mu^2\left((-1)^s(ul)-(ur)\right)}
{\sqrt{(ul)^2+\mu^2}\sqrt{(ur)^2+\mu^2}}\\
&+&i\varepsilon_{\mu\nu\alpha\beta}
\left[\frac{\mu^2\left((-1)^s r^{\alpha}-l^{\alpha}\right)u^{\beta}
+(ur)r^{\alpha}l^{\beta}}{\sqrt{(ur)^2+\mu^2}}\right.\\
&&\mbox{}\quad\left.-\frac{\mu^2\left((-1)^s l^{\alpha}
-r^{\alpha}\right)u^{\beta}
-(ul)r^{\alpha}l^{\beta}}{\sqrt{(ul)^2+\mu^2}}\right]\,.
\end{eqnarray*}

\section{Amplitudes of processes contributing to the radiative 
decay}\label{app:C}
Amplitudes of processes contributing to the radiative decay described by 
diagrams on Fig.~\ref{graph:3} (a)--(f) read
\begin{equation}
\Omega_{i}^{\mu}=\bar{v}(p)M_{i}^{\mu}v(k)
\end{equation}
where $i=a,b,c,d,e,f$ and:
\begin{eqnarray}
\lefteqn{M_{a}^{\mu}=\frac{g^2e}{2}\int\frac{d^4r}{(2\pi)^4}
\frac{1+\gamma^5}{2}\gamma^\nu}\nonumber\\
&\times&\frac{[(p-r)\gamma+m]\gamma^\alpha[(k-r)\gamma+m]}
{[r^2-m_{W}^{2}][(p-r)^2-m^2][(k-r)^2-m^2]}\nonumber\\
&\times&\gamma_\nu\frac{1-\gamma^5}{2}\,,
\end{eqnarray}
\begin{eqnarray}
\lefteqn{M_{b}^{\mu}=-\frac{g^2e}{2}\int\frac{d^4r}{(2\pi)^4}
\frac{1+\gamma^5}{2}\gamma^\beta}\nonumber\\
&\times&\frac{[r\gamma+m]}
{[r^2-m^{2}][(p-r)^2-m_{W}^{2}][(k-r)^2-m_{W}^{2}]} \nonumber\\
&\times&\left[(2p-k-r)_\lambda g^{\mu}_{~\beta}
+(2r-k-p)^\mu g_{\lambda\beta} \right.\nonumber\\
&&\mbox{}\quad\left.+(2k-r-p)_\beta g^{\mu}_{~\lambda}\right]
\gamma^\lambda\frac{1-\gamma^5}{2}\,,
\end{eqnarray}
\begin{eqnarray}
\lefteqn{M_{c}^{\mu}=
-\frac{g^2e}{8m_{W}^{2}}\int\frac{d^4r}{(2\pi)^4}
\left[(m-\kappa)I+(m+\kappa)\gamma^5\right]}\nonumber\\
&\times&\frac{[(p-r)\gamma+m]\gamma^\mu[(k-r)\gamma+m]}
{[r^2-m_{W}^{2}][(p-r)^2-m^{2}][(k-r)^2-m^{2}]}\nonumber\\
&\times&\left[(m-\kappa)I-(m+\kappa)\gamma^5\right]\,,
\end{eqnarray}
\begin{eqnarray}
\lefteqn{M_{d}^{\mu}=\frac{g^2e}{8m_{W}^{2}}\int\frac{d^4r}{(2\pi)^4}
\left[(m-\kappa)I+(m+\kappa)\gamma^5\right]}\nonumber\\
&\times&\frac{[(p-r)\gamma+m](k+p-2r)^\mu[(k-r)\gamma+m]}
{[r^2-m^{2}][(p-r)^2-m_{W}^{2}][(k-r)^2-m_{W}^{2}]}\nonumber\\
&\times&\left[(m-\kappa)I-(m+\kappa)\gamma^5\right]\,,
\end{eqnarray}
\begin{eqnarray}
\lefteqn{M_{e}^{\mu}=\frac{g^2e}{8}\int\frac{d^4r}{(2\pi)^4}
\gamma^\mu(1-\gamma^5)}\nonumber\\
&\times&\frac{[r\gamma+m]}
{[r^2-m^{2}][(p-r)^2-m_{W}^{2}][(k-r)^2-m_{W}^{2}]}\nonumber\\
&\times&\left[(m-\kappa)I-(m+\kappa)\gamma^5\right]\,,
\end{eqnarray}
\begin{eqnarray}
\lefteqn{M_{f}^{\mu}=\frac{g^2e}{8}\int\frac{d^4r}{(2\pi)^4}
\left[(m-\kappa)I+(m+\kappa) \gamma^5\right]}\nonumber\\
&\times&\frac{[r\gamma+m]}
{[r^2-m^{2}][(p-r)^2-m_{W}^{2}][(k-r)^2-m_{W}^{2}]}\nonumber\\
&\times&(1+\gamma^5)\gamma^\mu\,.
\end{eqnarray}
After calculations and renormalization, the integrals contributing to the 
form factor $G_V$ in the Eq.~(\ref{GV}) read 
\begin{eqnarray*}
I_a & =& -2\int_{0}^{1}d x \frac{x(x^2-3x+2)}
{x^2\kappa^2+x(m_{W}^{2}-\kappa^2-m^2)-m_{W}^{2}}\\
& \simeq& \frac{4}{3m_{W}^{2}}\,,
\end{eqnarray*}
\begin{eqnarray*}
I_b & =& \int_{0}^{1} d x \frac{x^2(2x+1)}
{x^2\kappa^2+x(m^2-\kappa^2-m_{W}^{2})-m^2}\\
& \simeq& -\frac{7}{6m_{W}^{2}}\,,
\end{eqnarray*}
\begin{eqnarray*}
I_c & =&   \frac{1}{m_{W}^{2}}
\int_{0}^{1}d x \frac{x^2[x(\kappa^2-m^2)-(m^2+\kappa^2)]}
{x^2\kappa^2+x(m_{W}^{2}-\kappa^2-m^2)-m_{W}^{2}}\\
& \simeq& \frac{2m^2}{m_{W}^{4}}\ln{\left(\frac{m_W}{m}\right)^2}\,,
\end{eqnarray*}
\begin{eqnarray*}
I_d & =&  \frac{1}{m_{W}^{2}}
\int_{0}^{1}d x \frac{x(x-1)[x(\kappa^2-m^2)+2m^2]}
{x^2\kappa^2+x(m^2-\kappa^2-m_{W}^{2})-m^2}\\
& \simeq& \frac{5m^2+\kappa^2}{6m_{W}^{4}}\,,
\end{eqnarray*}
\begin{eqnarray*}
I_e+I_f & =& -\int_{0}^{1}d x \frac{x^2}
{x^2\kappa^2+x(m^2-\kappa^2-m_{W}^{2})-m^2}\\
& \simeq& \frac{1}{2m_{W}^{2}}\,.
\end{eqnarray*}

\section{Sum over photon polarization vectors}\label{app:D} Sum over photon 
polarization vectors (on the mass shell $q^2=0$) in the CT synchronization 
reads:
\begin{eqnarray}
 \Pi^{\mu\nu}(q) &=& 
 \sum_{\begin{array}{c}\scriptstyle \lambda=-1 \\
\scriptstyle \lambda\ne 0\end{array}}^1
 e^{\mu}_{~\lambda}e^{\nu}_{~\lambda}\nonumber\\
&=& - \frac{1}{(uq)^2}\left[(uq)^2g^{\mu\nu}(u) +
 q^\mu q^\nu \right.\nonumber\\
&&\left.- (uq)(u^\mu q^\nu+u^\nu q^\mu)\right]\,.
 \label{tensorpol}
 \end{eqnarray}
Notice, that $q_\mu\Pi^{\mu\nu}=u_\mu\Pi^{\mu\nu}=0$, so
$e^{\mu}_{~\lambda}$, $q^\mu$ and $u^\mu$ form an pseudo-orthogonal basis.

\end{document}